\documentclass[11pt,twoside]{article}

\usepackage{asp2006}
\usepackage{epsf}
\usepackage{psfig}
\usepackage{lscape}
\usepackage{graphicx}

\begin{document}
\title{An AKARI-FIS survey of post-AGB stars and (proto) planetary nebulae:
an analysis of extended emission and the spectral energy distribution.} 

\author{Nick Cox,$^1$ Pedro Garc\'ia-Lario,$^1$ Anibal Garc\'ia-Hern\'andez,$^2$ Arturo Manchado$^{2,3}$}
\affil{%
$^1$ Herschel Science Centre, European Space Astronomy Centre, European Space
Agency, Spain\\
$^2$ Instituto Astrof\'isica de Canarias, Spain\\
$^3$ CSIC, Spain} 

\begin{abstract}
We present first preliminary results from AKARI/FIS pointed observations of post-AGB stars and planetary nebulae (PNe). 
A first analysis of the radial (azimuthally averaged) profile of the observed sources shows no evidence for excess
emission due to the presence of circumstellar dust. No (detached) circumstellar faint dust-shells are seen in the image maps. 
Also, we present here first results of aperture flux photometry at wavelengths of 65, 90, 140 and 160~$\mu$m. 
Results are compared with IRAS flux densities as well as the $\beta$ release of the FIS Bright Source Catalog. 
Finally, spectral energy distributions  
are given, by way of an example, for two individual targets in our sample.
\end{abstract}

\section{Introduction}
The stellar evolution from the asymptotic giant branch (AGB) phase into planetary nebulae (PNe) is
still shrouded in mystery. In particular, the mass loss mechanisms during the AGB and post-AGB phases
are essential ingredients for stellar evolution models.
It is therefore important to recover the mass loss history experienced by stars in their transition from 
AGB to PNe stage. 

AGB stars have high mass loss rates of up to 10$^{-5}$ M$_{\odot}$/year, but the mechanism and evolution of mass loss rates
are poorly understood. The mass loss drops to 10$^{-7}$ M$_{\odot}$/year when the star evolves to the post-AGB / proto-PN phases.
Spherical circumstellar (CS) shells continue to drift away, pushed by the post-AGB wind and, somehow, transformed into aspherical PNe
with fast outflows. For young PNe and proto-PNe the outer shells are not yet disrupted by fast post-AGB
winds and therefore a fossil record of the original AGB mass loss may still be preserved.
The distribution of cool dust in the extended envelopes of post-AGBs and PNe can be studied by measuring their
infrared properties (e.g. size and flux).
The structure of extended dust emission depends on the mass-loss mechanism (e.g. constant rate or fluctuations).
The details are still uncertain; earlier ISOPHOT observations of two post-AGBs suggested episodic mass loss (Speck et al. 2000), 
while recent Spitzer observation revealed no CS shells in one of these sources (Do et al. 2007).

\section{Observations} 
During the AKARI (\citealt{murakami07}) cold phase we obtained with
the Far Infrared Surveyor (FIS; \citealt{kawada07}) instrument  images of 13 post-AGB stars and PNe in the four 
far-infrared (FIR) filter bands at 60, 90, 140 and 160~$\mu$m.

The FIS Slow Scan Toolkit (ver. 20070914; Verdugo et al. 2007), implemented in IDL, was used to perform the FIS data 
reduction and to produce the maps.
We applied the usual bad-pixel correction, stray-light removal, median \& boxcar filtering.
For bright point sources there is appreciable cross-talk, up to 10\%, (only in the SW detector) as well
as ghosts (at fixed positions in both SW and LW detectors).
The final calibrated maps are in units of MJy/sr.

\begin{figure}[th!]
\begin{center}
   \includegraphics[width=1\textwidth]{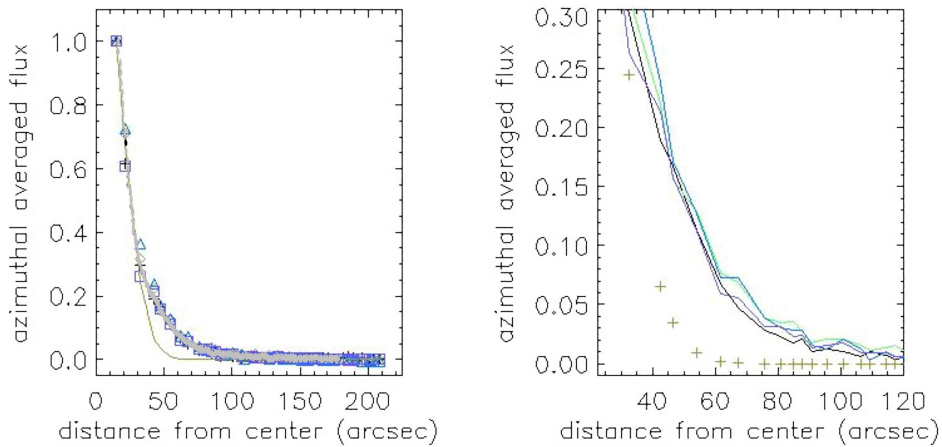}
\end{center}
\caption{The azimuthally averaged radial profiles normalized to peak flux (for the 90~$\mu$m SW band) 
are shown for four targets in our sample.
In the left panel the symbols show the radial profiles of our FIS targets and the solid thin line gives the Gaussian fit. 
The departure from Gaussianity can be clearly seen around 60'' from the peak.
The instrumental PSF for the the 90~$\mu$m SW band (extracted from the FIS Data User Manual) is over plotted in grey.
The latter is traced closely by the observed radial profiles.
In the right panel we show this in more detail (solid lines are targets and symbols indicate the Gaussian fit)
of the point source PSF.
}
\label{fig:profile}
\end{figure}

\section{Results I: Radial Profiles \& Extended Emission}
We derived the radial (azimuthally averaged) profile for thirteen post-AGB stars and (P)PNe, which we
subsequently compare to each other and to the point-spread function (PSF; Figure~\ref{fig:profile}).
The PSF of Ceres has a reported FWHM of 37'' (65~$\mu$m), 39'' (90~$\mu$m), 58'' (140~$\mu$m) and 61'' (160~$\mu$m) for the four bands, respectively.
At low flux levels the instrumental PSF deviates from a Gaussian profile.

\begin{figure}[ht!]
\begin{center}
  \includegraphics[width=0.75\textwidth,height=6.75cm]{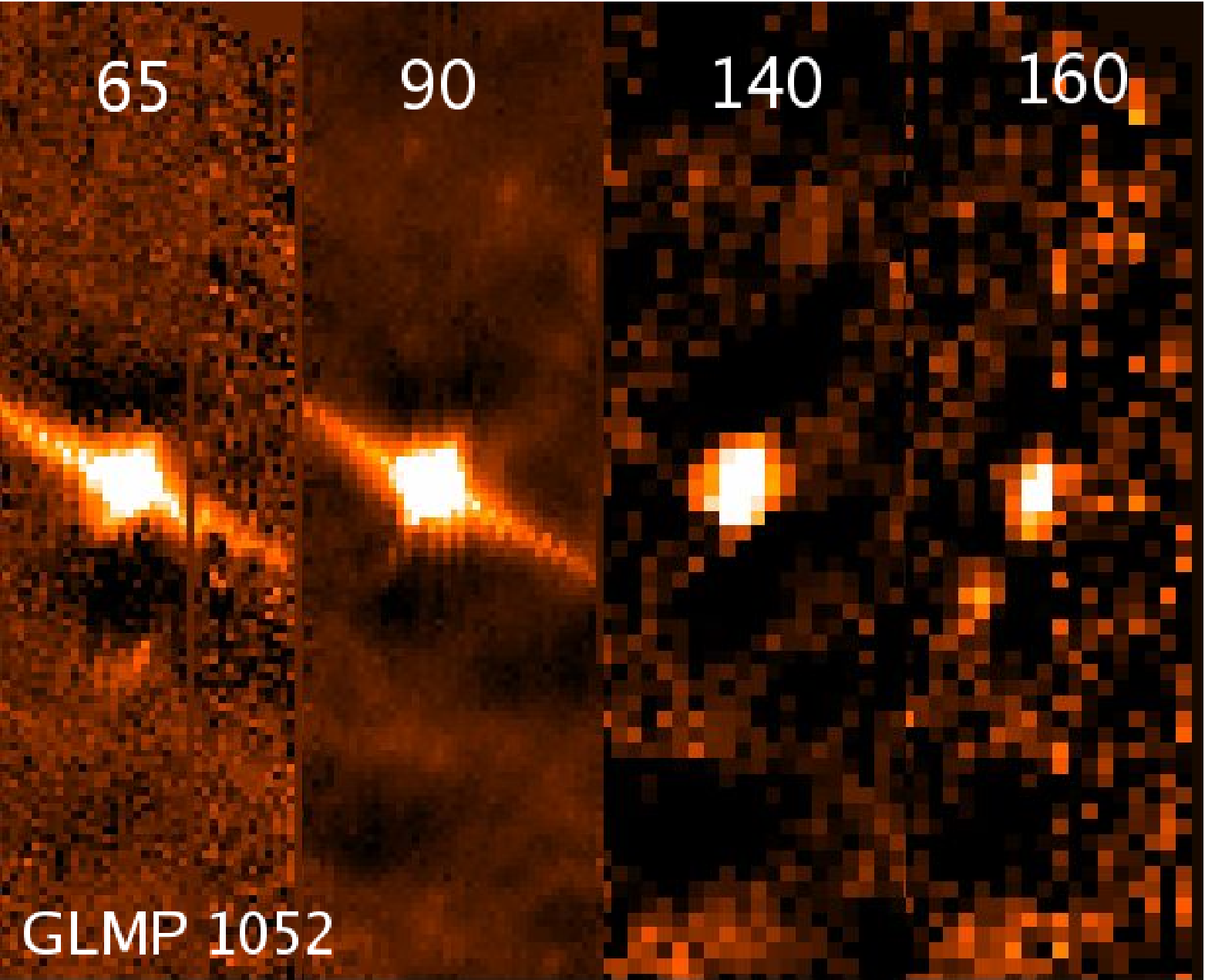}
  \includegraphics[width=0.75\textwidth,height=6.25cm]{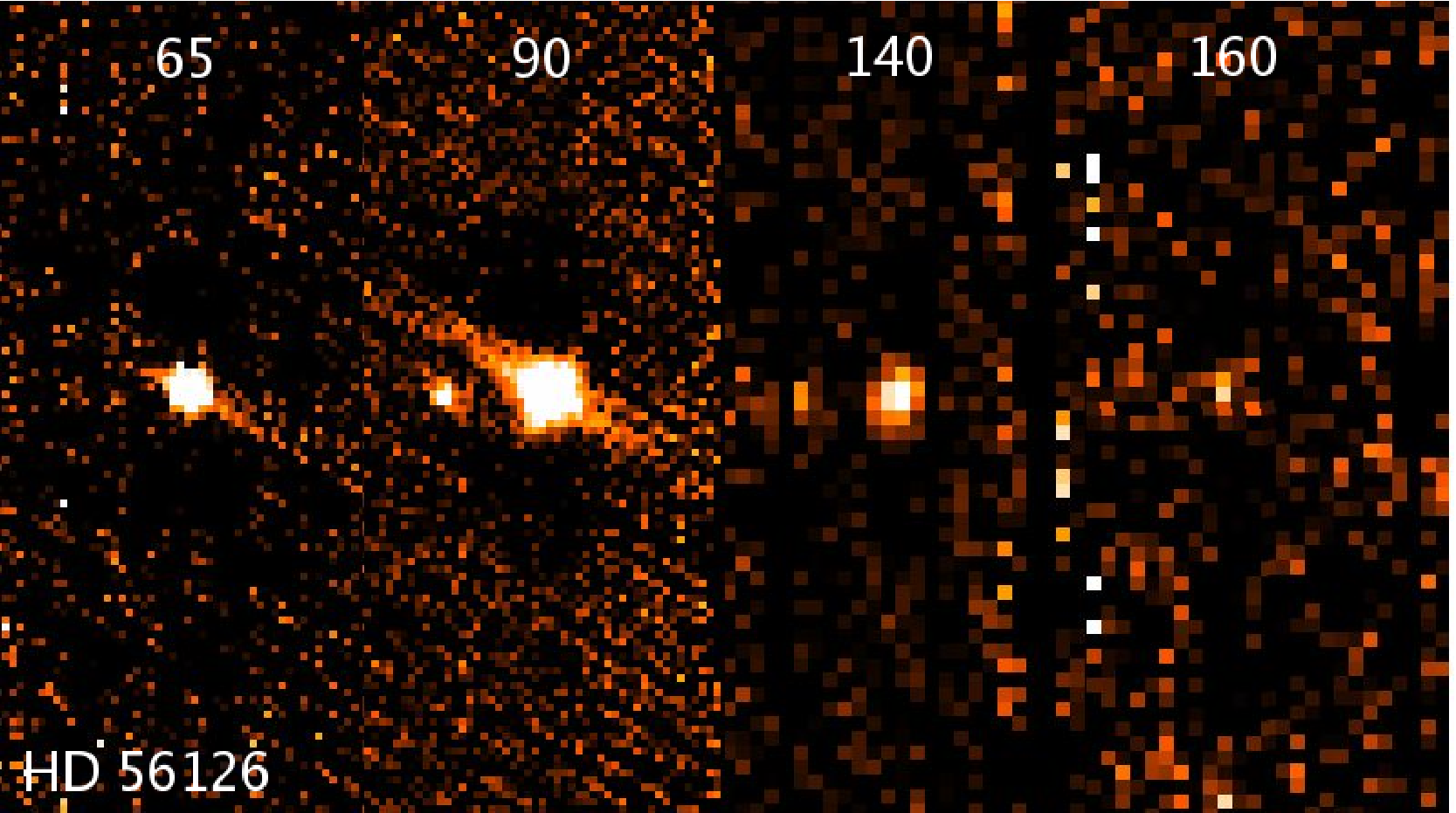}
\end{center}
\caption{Photometric image maps (10 x 40 arcmin; in MJy/sr) are shown for GLMP\,1052 (top) and HD\,56126 (bottom) at 65, 90, 140 and 160~$\mu$m.
Logaritmic brightness levels are adjusted for each of the individual panels to provide optimal contrast.
}
\label{fig:maps}
\end{figure}

For the SW bands at 65 and 90~$\mu$m there is no evidence for the presence of circumstellar dust-shells
connected directly to the (P)PN. In other words, all derived radial profiles are consistent with the PSF
derived from know point sources (such as Ceres).
The long wavelength images at 140 and 160~$\mu$m show structured emission - due to cirrus - at and near 
the target positions; only in a few cases are the point sources detected at these far IR wavelengths.
CS dust shells, if present, are very faint or wide and disappeared in the background.
For illustration, image maps of two targets, GLMP\,1052 and HD\,56126, are shown for each of the four filter bands (Figure~\ref{fig:maps}).

\begin{figure}[ht!]
\begin{center}
   \includegraphics[angle=-90,width=0.975\textwidth]{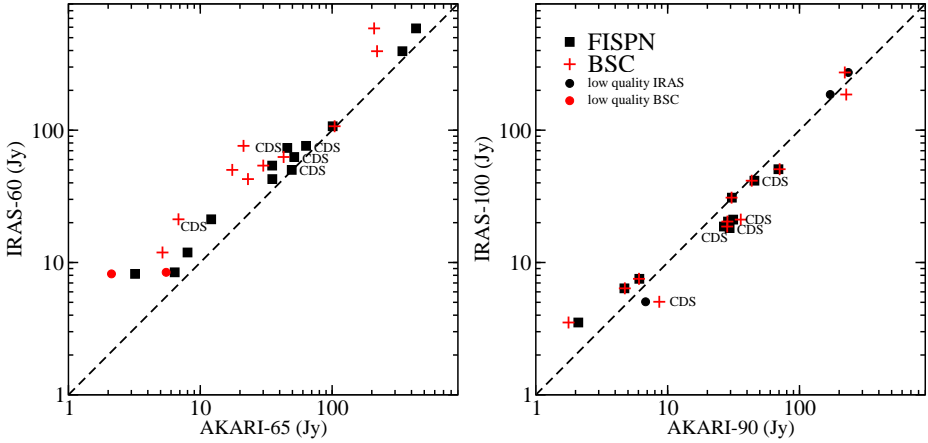}
\end{center}
\caption{Comparison of AKARI and IRAS flux densities (FIS pointed observations 
as black squares/circles and FIS BSC as red crosses/circles.}
\label{fig:flux_sw}
\end{figure}

\begin{figure}[ht!]
\begin{center}
  \includegraphics[angle=-90,width=0.975\textwidth]{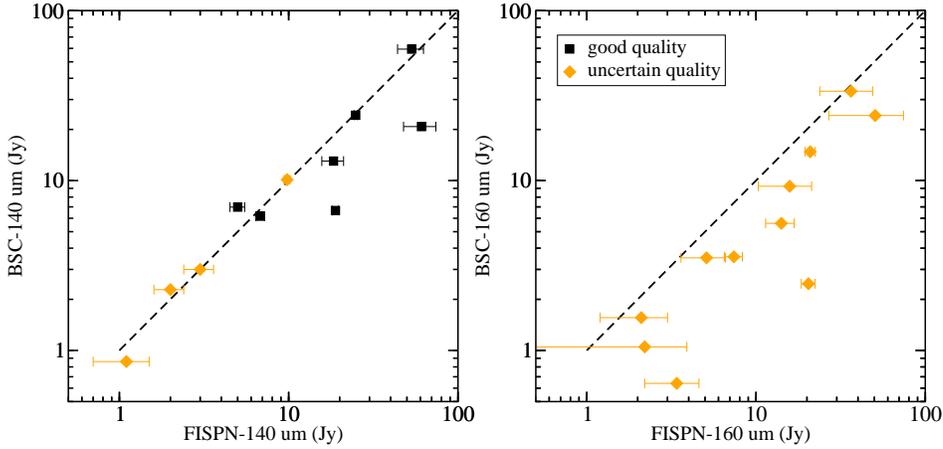}
\end{center}
\caption{Comparison between AKARI flux from pointed FIS observations and from the FIS BSC. 
Squares correspond to good quality BSC data and diamonds are of uncertain quality in the BSC.}
\label{fig:flux_lw}
\end{figure}

\section{Results II: Aperture Flux Photometry}
We performed standard aperture flux photometry on the calibrated maps.
When applicable we apply an aperture correction and a point-to-diffuse correction factor
(for details we refer to the FIS Slow Scan Toolkit). We did not apply a colour correction at this point.
PNe are strong IR emitters and the nearby extended envelope - if present - contributes negligible (within error bars) to the overall flux.

We compare the 65 and 90~$\mu$m fluxes of the pointed observations both with the IRAS 60 and 100~$\mu$m 
and FIS-BSC 65 and 90~$\mu$m fluxes in Figure~\ref{fig:flux_sw} (left and right panel respectively). 
The flux densities reported in the current $\beta$-1 release of the FIS BSC are under-estimated with respect to
both the IRAS and FIS pointed observations (this work).
Note that IRAS fluxes at 60~$\mu$m are indeed slightly lower than AKARI fluxes at 65~$\mu$m, 
as is indeed expected for dust temperatures between 20-70~K.
In Figure~\ref{fig:flux_lw} we compare the pointed observations with the reported fluxes (140 and 160~$\mu$m) in the
all-sky FIS survey.  
Coordinates of IRAS and AKARI sources (both pointed and survey) agree all within 15''. 
Note that one source in our sample (IRAS\,19500-1709) is not in the BSC.

\section{Results III: Spectral Energy Distribution}
We use the new results of the aperture flux photometry to construct spectral energy distributions (SEDs).
Previous data, from 2MASS, MSX, IRAS and ISO, were obtained via the Toru\'n catalogue of Galactic post-AGB and related objects
\citep{szczerba09}.

\begin{figure}[t!]
\begin{center}
  \includegraphics[angle=-90,width=.9\textwidth]{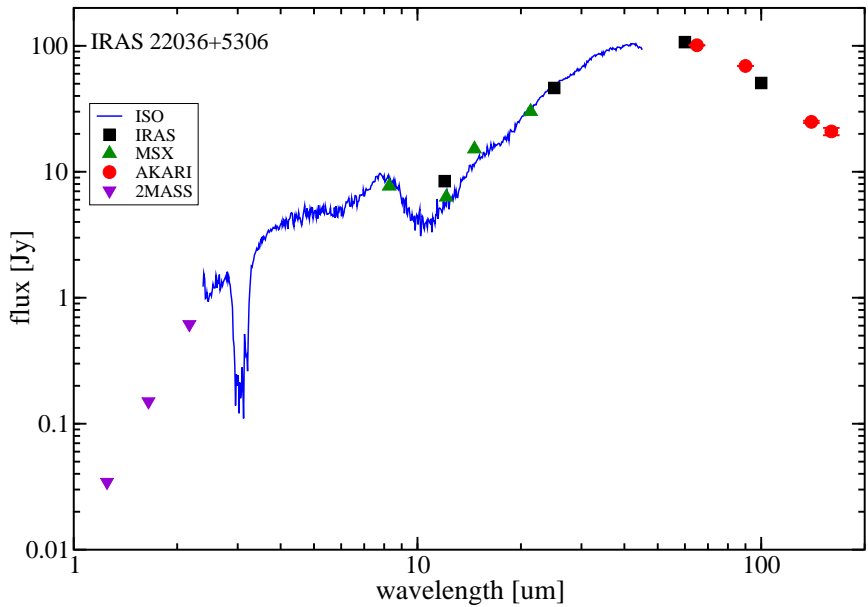}
\end{center}
\caption{Spectral energy distribution of GLMP\,1052 (IRAS\,22036+5306)}
\label{fig:sed1}
\end{figure}

\begin{figure}[h!]
\begin{center}
  \includegraphics[angle=-90,width=.9\textwidth]{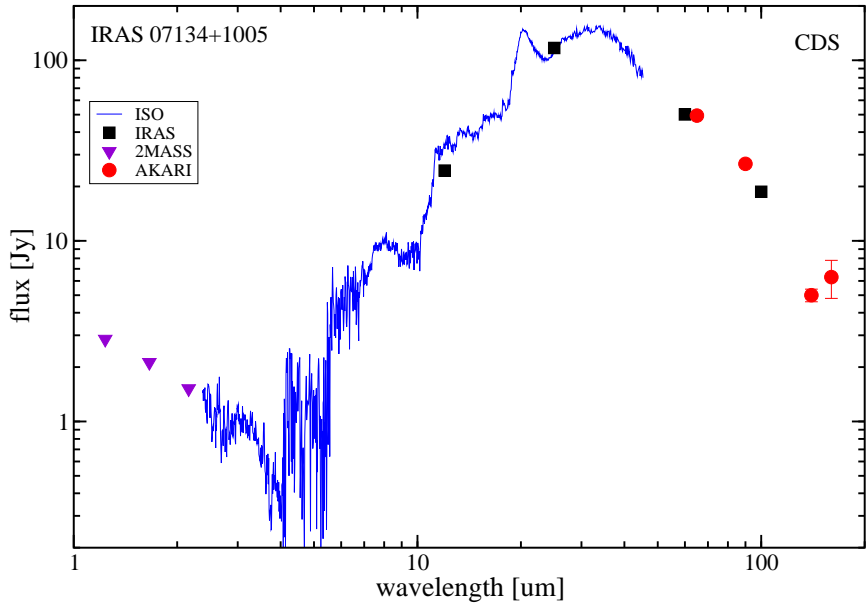}
\end{center}
\caption{Spectral energy distribution of HD\,56126 (IRAS\,07134+1005)}
\label{fig:sed2}
\end{figure}

In Figures~\ref{fig:sed1} and~\ref{fig:sed2} we show the SEDs for two sources, 
GLMP\,1052 and HD\,56126, in our sample.
The AKARI data extend the SED into the FIR ($>$100~$\mu$m) improving
the constraints necessary to fit dust emission models.

\section{Summary and Outlook}

Additional analysis of the radial profiles in 3D (e.g. differences in cross-scan
and in-scan profiles) are in progress to elucidate the possible presence
of very compact CS dust close to the star.
Future observations, with for example Herschel, may shed more light on these
elusive dust shells.

\acknowledgements 
We thank the organizers for their hospitality and excellent, efficient meeting.
This research is based on observations with AKARI, a JAXA project with the participation of ESA.

\end{document}